\documentclass[12pt]{article}

\usepackage{amsfonts,amsmath,amssymb,color,subfigure}
\usepackage{enumerate}
\usepackage{hyperref}
\usepackage{bbm}
\usepackage{nicefrac}
\usepackage[all]{xy}
\usepackage{graphicx}

\usepackage{booktabs}
\newcommand{\ra}[1]{\renewcommand{\arraystretch}{#1}}

\newcommand{\be}{\begin{equation}}
\newcommand{\ee}{\end{equation}}

\newcommand{\bea}{\begin{eqnarray}}
\newcommand{\eea}{\end{eqnarray}}

\newcommand{\bes}{\begin{subequations}}
\newcommand{\ees}{\end{subequations}}

\usepackage{multirow}
\usepackage{rotating}

\let\OLDthebibliography\thebibliography
\renewcommand\thebibliography[1]{
  \OLDthebibliography{#1}
  \setlength{\parskip}{3.5pt}
}

\begin{document}

\makeatletter
\renewcommand{\theequation}{\thesection.\arabic{equation}}
\@addtoreset{equation}{section}
\makeatother

\begin{titlepage}

\begin{flushright}
\end{flushright}

\vspace{1cm}

\begin{center}
   \baselineskip=16pt
   \begin{Large}\textbf{
Quasinormal ringing on the brane}
   \end{Large}
   		
\vspace{25pt}
		
Hyeyoun Chung, Lisa Randall, Maria J. Rodriguez and  Oscar Varela
		
\vspace{25pt}

	\begin{small}

	{\it Center for the Fundamental Laws of Nature,\\
	Harvard University, Cambridge, MA 02138, USA }
		
	\end{small}

\vskip 50pt

\end{center}

\begin{center}
\textbf{Abstract}
\end{center}

\begin{quote}

While the linear behavior of gravity in braneworld models is well understood, much less is known about full non-linear gravitational effects. Even when they agree at the linear level, these could be expected to distinguish braneworlds from a lower-dimensional theory with no brane. Black holes are a good testing ground for such studies, as they are nonlinear solutions that would be expected to reflect the background geometry.  In particular, we assess the role of black hole quasinormal modes in gravitational experiments devised to be sensitive to the existence of the brane, in a lower-dimensional setting where we have analytical control. We compute quasinormal modes of brane-localized black holes and find that they follow the entropy of the corresponding black hole. This observation allows us to conclude that, surprisingly, the scattering problem we consider, at least in some regimes,  does not distinguish between non-linear gravitational effects of black holes in AdS space with a brane and  black holes in a spacetime of one lower dimension.

\end{quote}

\vfill

\end{titlepage}

\tableofcontents


\section{Introduction}

The Randall-Sundrum (RS) braneworld models \cite{RandallSundrumI, RandallSundrumII} were originally introduced as a potential solution to the hierarchy problem. They consist of five-dimensional anti-de-Sitter (AdS) spacetime with either one (the RSII model) or two (the RSI model) Minkowski branes embedded within them. These models differ from the traditional Kaluza-Klein approach to extra dimensions in that the single extra dimension is warped and, in the RSII case, non-compact. The Standard Model fields live on the four-dimensional brane while gravity propagates also along the extra fifth dimension, though it is mostly localized on and near the brane due to the bulk AdS curvature. Since its original discovery, the applicability of the RS models  has expanded far beyond its original roots. 

Gravitational effects in RSII models have been well studied at the linearized level. In this approximation, four-dimensional linearized Einstein gravity on the brane is reproduced up to the AdS scale \cite{RSLinGrav1, RSLinGrav2, RSLinGrav3}. Beyond the linear regime, however, much less is known. In fact controversy persisted for some time over whether static black hole solutions even exist in the RSII scneario. Investigating the fully-fledged, non-linear behavior of gravity in RS models is crucial in order to find potential phenomenological signatures of whether or not we live in a braneworld universe. It is thus important to devise physical observables that allow us to determine how gravitational systems behave with and without warped extra dimensions, and thus let us distinguish between a braneworld scenario and a case with no extra dimensions other than the regular four. In this paper, we explore the role, in a simplified lower-dimensional setting, of black hole quasinormal modes (QNMs) as such potentially distinguishing physical observables.

Black hole backgrounds provide an ideal setup to study gravity in RS models beyond the linearized approximation, as they involve fully backreacted gravitational effects. See \cite{Gregory:2008rf, Tanahashi:2011xx} for some reviews. However, black holes localized on the four-dimensional RSII  brane were conjectured in \cite{Tanaka, EmparanConj} not to exist. It was later argued \cite{Fitzpatrick}  that the existence of static, localized braneworld black hole solutions is in principle not ruled out by the arguments of \cite{Tanaka, EmparanConj}, as these rely on an extrapolation of weak coupling calculations to the strongly interacting regime --something not necessarily justified. In fact, two groups have now found numerical solutions for RSII braneworld black holes using independent methods \cite{Wiseman, Page}, although it is not yet known whether these solutions are dynamically stable. If the solutions are unstable, then it is still possible that the classical dynamical instability in the bulk could correspond to phenomenological effects on the brane that would distinguish the five-dimensional braneworld black hole from its counterpart in a strictly four-dimensional universe.

In any case, while the construction of these black hole solutions is a remarkable result, the fact that the solutions are  known only numerically somewhat limits their potential for further investigations. For this reason, we turn to the simplified scenario provided by the lower-dimensional version of RS with a three-dimensional brane embedded in a four-dimensional AdS bulk. In this setting, explicit, analytical solutions corresponding to brane-localized black holes are known, both for flat \cite{Emparan:1999wa} and curved \cite{Emparan:1999fd} three-branes within the AdS four-dimensional bulk. We will be more interested in the latter case,  \cite{Emparan:1999fd},  with negative curvature on the brane, since only in this case a counterpart of the brane-localized black hole exists in a strictly three-dimensional universe with no brane: the Ba\~nados-Teitelboim-Zanelli (BTZ) black hole \cite{Banados:1992wn}. 

We can thus compare suitable characteristics of these braneworld black holes to those of the strictly three-dimensional BTZ solution. Specifically, we study whether an asymptotic observer in three-dimensional AdS space who sent waves to a black hole and measured the corresponding QNMs could decide on the existence of a fourth dimension. Namely, we address the question of whether QNMs are different for the black holes in the two different geometries. We will argue that, remarkably, this experiment will in general not distinguish the two scenarios, at least for large black holes. This is surprising since such experiments clearly probe the nonlinear regime.

Recall that the QNMs of a black hole are its characteristic modes of vibration when perturbed with fully ingoing waves at the horizon. The boundary value problem for fields on black hole backgrounds is dissipative, which leads to an imaginary part for the characteristic vibrational frequencies. If, in the usual conventions (see equation (\ref{eq:scalarconfined}) below), this imaginary part is negative, then the system exponentially decays in time to its unperturbed, ground state. In this case, the first few QNMs thus determine the `ring down' of the perturbation. This is the situation we find for the black holes under consideration in this paper. Instead, positive imaginary parts for these frequencies would signal instabilities of the type, for example, of those that holographically produce a superconducting phase transition \cite{Hartnoll:2008vx}. The QNMs of the BTZ black hole have been computed for various perturbations coupled  minimally \cite{Cardoso:2001hn, Birmingham:2001pj} and conformally \cite{Chan:1996yk} to the black hole background. QNMs for various constructions of black holes on branes, different from our present setting, have been obtained in \cite{Kanti:2005xa,Kanti:2006ua,Nozawa:2008wf}. References \cite{Aros:2002te,Oliva:2010xn} study the QNMs of some topological black holes that bear some resemblance to the brane-localized black holes that we consider in this paper. See also \cite{Motl:2002hd,Motl:2003cd,Castro:2013lba,Castro:2013kea} for the calculation of QNMs using monodromy methods, \cite{Gonzalez:2010vv,Cardoso:2006nh,Cvetic:2014ina} for other related work and, more generally, \cite{QNMRev1,Konoplya:2011qq} for recent reviews.

We are mostly interested in the QNMs of perturbations with a profile entirely on the brane, as this case is amenable to comparison with the braneless BTZ case. However, we will also study bulk-probing perturbations for reasons to be discussed below. For simplicity, we focus in this paper on scalar perturbations. Our calculations show that the (imaginary part of the) QNMs for scalar brane-confined perturbations track down the entropy of the brane-localized black holes, exactly as for the pure BTZ black hole \cite{Cardoso:2001hn}. In other words, the measurement of QNMs of scalar perturbations turns out to be equivalent to the measurement of the entropy of the relevant black hole, both in the pure BTZ case and in the brane-localized black hole  case.

The outcome of the scattering experiment will thus critically depend on which black hole dominates the phase diagram. The known phase diagram \cite{Emparan:1999fd} includes a `BTZ black string' (namely, the warped product of the BTZ black hole with the extra dimension) and, for small black hole mass, two further branches of brane-localized black holes. Some of these black hole branches will typically be unstable to decay, by emission of droplets into the bulk, into stable `multicenter' configurations composed of a localized black hole together with the droplets. By the arguments of \cite{Fitzpatrick}, such multicenter configurations will have an entropy very close to the original black hole in the phase diagram of  \cite{Emparan:1999fd} they decayed from. 

For large, `astrophysical' black holes\footnote{In quotes, only because these are black holes in three-dimensional AdS.}, the dominant black hole would correspond to either the BTZ string or a closely related multicenter phase with essentially the same entropy and, thus, essentially the same QNMs. Scattering experiments will therefore be insensitive to the presence of the brane. For small black hole mass, when according to the classical calculation all three known branches exist, the most entropic branch would in principle, but not necessarily, be the dominant phase. In order to check for instabilities that could potentially decide which phase is stable, we consider a different type of scalar perturbation: one that is now also allowed to probe the extra fourth dimension. In this case we also find that all branches are stable under this type of perturbations. This is not an exhaustive analysis of all possible perturbations, so this analysis is inconclusive.  In any case, a fully quantum analysis might be required to reliably determine small stable black holes. Nonetheless, we will proceed as if all branches --small and large-- are relevant and check how well quasinormal frequencies agree in the braneworld and lower-dimensional scenarios. We will nevertheless still argue that, also for small black holes, it might be difficult for this experiment to distinguish a warped higher-dimensional theory from a lower-dimensional one at the full non-linear level.

The rest of the paper is organized as follows. Section \ref{sec:Setup} briefly reviews the braneworld black holes \cite{Emparan:1999fd} that we consider. In section \ref{sec:QNMBrane} we compute the QNMs of braneworld black holes created by brane-confined scalar perturbations. Stability against bulk-probing perturbations is tested in section \ref{sec:QNM}. Section \ref{sec:Discussion} discusses the relation of the QNMs created by brane-confined perturbations  of these brane-localized black holes to those \cite{Cardoso:2001hn} of the BTZ black hole, and the (in)ability of the present experiment to detect the brane. For the sake of comparison with our results we rederive, in appendix \ref{app:BTZ}, the BTZ QNMs \cite{Cardoso:2001hn} using the powerful monodromy methods of \cite{Castro:2013kea}. Finally, appendix \ref{app:dSSch} contains a technical note.

\section{Braneworld black holes}
\label{sec:Setup}

Black holes localized on the brane are necessarily accelerating, since they do not follow a geodesic in AdS. For this reason, the construction \cite{Emparan:1999fd} of analytic black holes localized on the three-dimensional brane starts from the four-dimensional AdS C-metric  \cite{Plebanski:1976gy}, which describes accelerating black holes in AdS. This metric reads
\begin{eqnarray} \label{AdSCmetric}
ds^2=\frac{1}{A^2(x-y)^2}\left(H(y) dt^2-\frac{dy^2}{H(y)}+\frac{dx^2}{G(x)}+G(x)d\phi^2\right) \; , 
\end{eqnarray}
where
\begin{eqnarray}
H(y)&=&-\lambda +y^2-2\mu\,A\,y^3 \,,\qquad \text{with} \,\,\, \lambda>0 \, , \label{eqH}\\
G(x)&=&1+x^2-2\mu\,A\,x^3 \, .\label{eqG}
\end{eqnarray}
For all values of the parameters $A \neq 0$, $\mu$ and $\lambda$, this is  a solution to the four-dimensional Einstein equation with negative cosmological constant, $R_{AB}=-(3\,l^{-2}_4)\,g_{AB}$,  $A,B=1, \ldots ,4$, with the four-dimensional AdS scale given by $l^{-2}_4 \equiv A^2(1+\lambda)$. We have also fixed a further constant, $k=1$, with respect to the solution presented in  \cite{Emparan:1999fd}. 

This metric was employed in \cite{Emparan:1999fd} to construct braneworld black holes with negative  (if $\lambda>0$)  cosmological constant on a three-brane inside the four-dimensional bulk. Their procedure in fact involves  two negatively curved branes, introduced by cutting the spacetime  (\ref{AdSCmetric}) along two hypersurfaces and then gluing at either side of each of them --see that reference for further details. The simplest choice  \cite{Emparan:1999fd} for those hyperurfaces corresponds to take them at $x=0$ and $y=0$ (see \cite{Anber:2008qu} for other choices).  Equivalently, defining\footnote{Under this coordinate transformation, the metrics (\ref{AdSCmetric}) are of the form
\begin{eqnarray}
ds^2=\frac{dr^2}{r^2 l^{-2}_4-\lambda}+r^2 A^2 g_{ab}dx^{a}dx^{b}
\end{eqnarray}
where $ds^2_b=g_{ab}dx^{a}dx^{b}$ is the induced metric on the 2-brane.} $r=\frac{\sqrt{y^2+\lambda x^2}}{A(x-y)}$, these hypersurfaces corresponds to the slices of constant $r=1/A$ and $r=\sqrt{\lambda}/A$. The resulting space describes a black hole localized on one of the branes (at $x=0$) on the patch $0\le x$ and $y \le 0$. The brane is negatively curved with an effective three-dimensional AdS scale $l_3^{-2}=\lambda\,A^2$. It is worth emphasizing that the parameters $A$ and $\lambda$ determine both cosmological constants $l_3$ and  $l_4$, on the brane and in the bulk, respectively. As we will now review, different ranges of the parameter $\mu$ correspond to different branches of braneworld black holes.

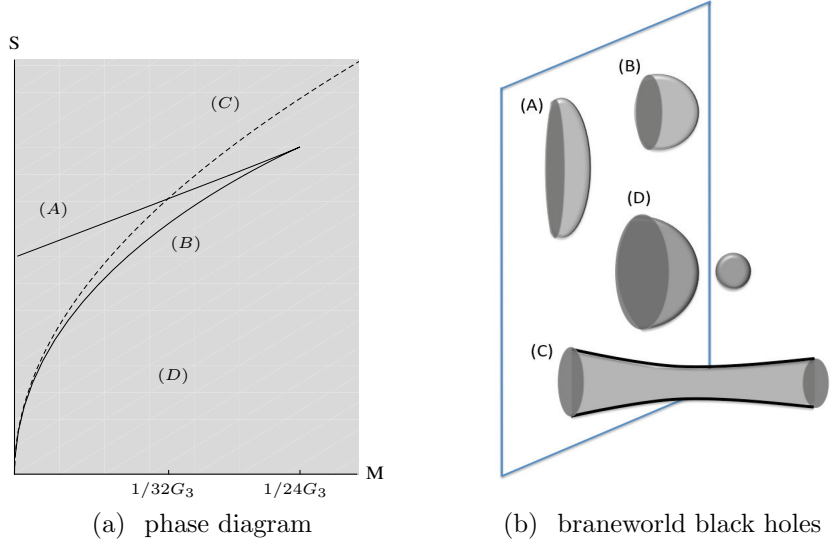
\begin{figure*}
    \centering 
\subfigure[\ phase diagram]{
    \includegraphics[width=5cm,height=6cm]{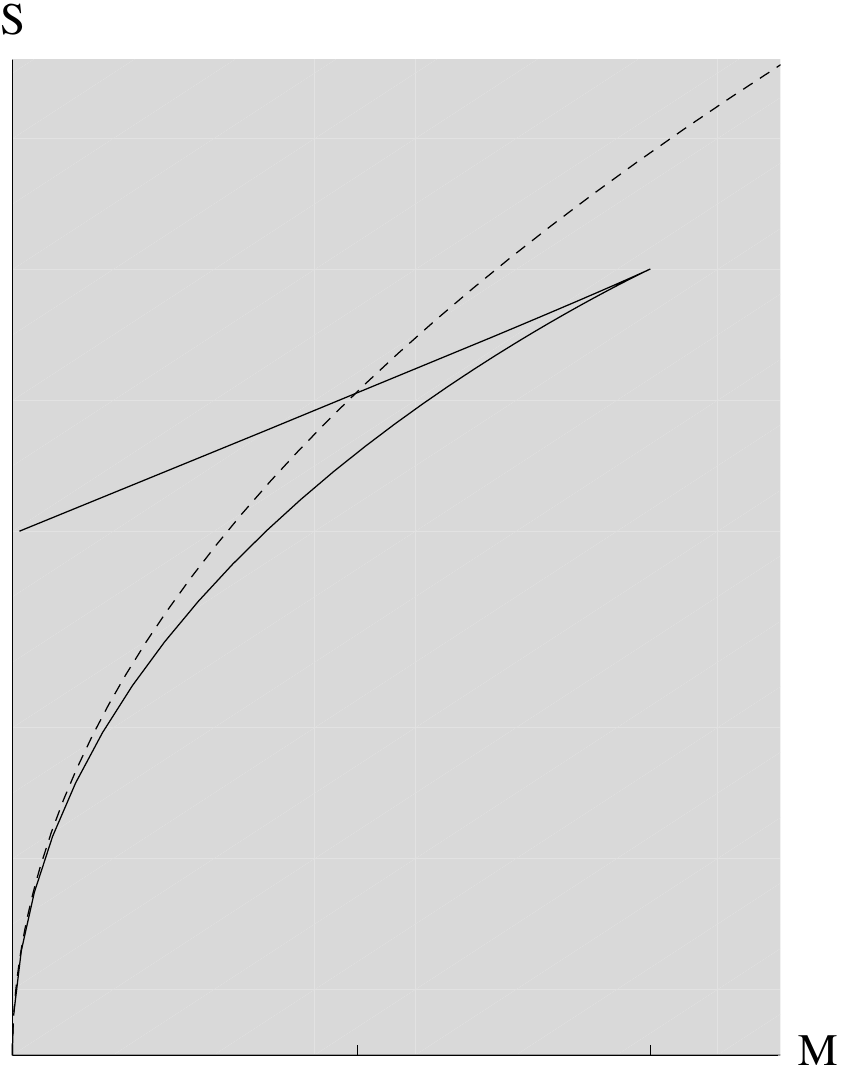}
    \begin{picture}(0,0)(0,0)
\put(-135,103){\tiny{$(A)$}}
\put(-85,90){\tiny{$(B)$}}
\put(-70,143){\tiny{$(C)$}}
\put(-90,40){\tiny{$(D)$}}
\put(-100,-3){\tiny{$1/32G_3$}}
\put(-50,-3){\tiny{$1/24G_3$}}
\end{picture}
 }   
    \hspace{0.5cm}
    \subfigure[\ braneworld black holes]{
      \includegraphics[width=5cm,height=6.5cm]{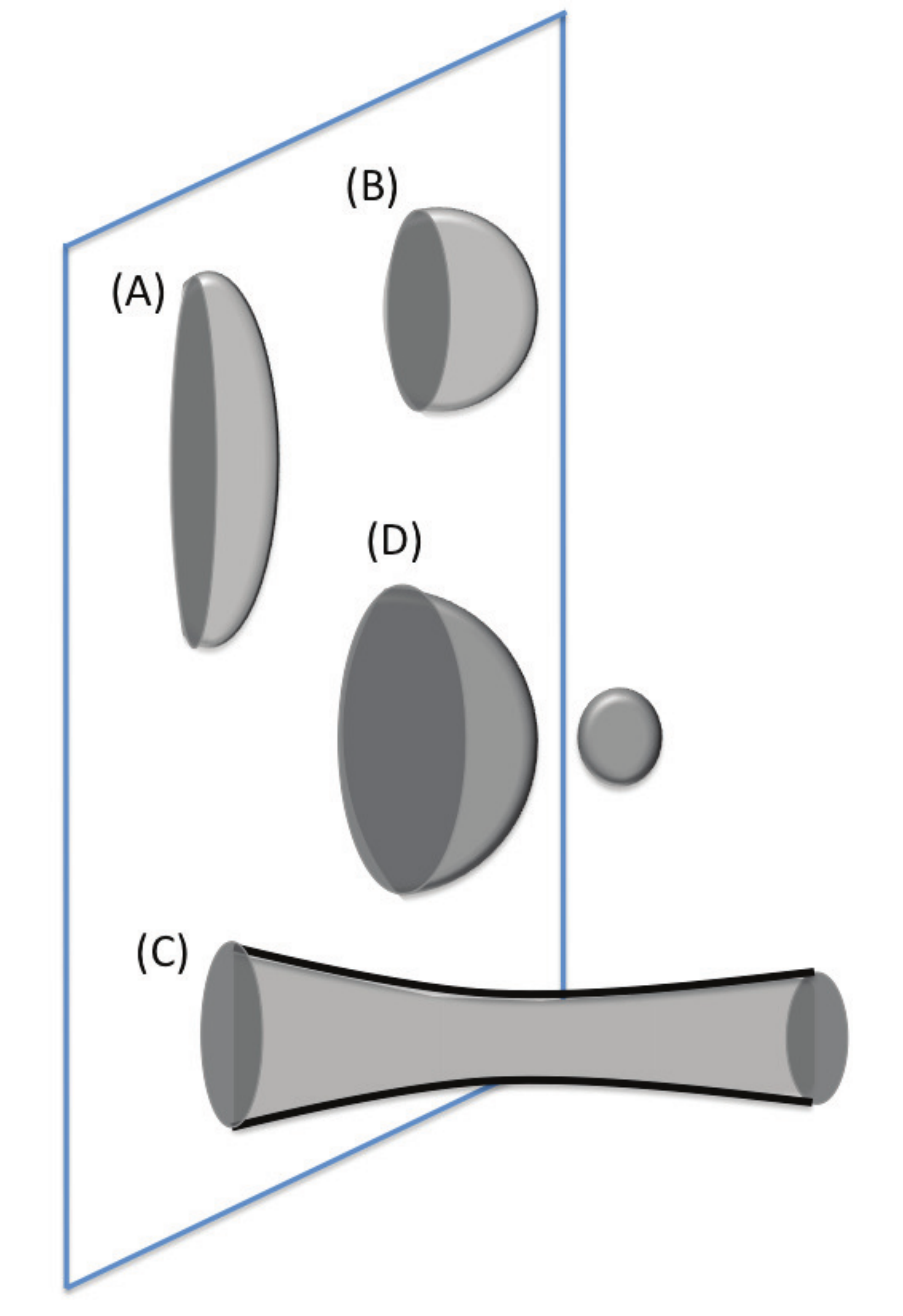}    
      }
    \caption{Entropy versus mass for the BTZ-black strings ({\it dashed line} $(C)$), localized BTZ-like black holes ({\it solid lines} $(A)$ and $(B)$) and conjectured phase of black hole-droplet configuration ({\it gray region} $(D)$).}\label{fig:phasediag}
\end{figure*}

Defining 
\begin{eqnarray} \label{eq:rhocoord}
\rho=-\frac{1}{y} \; ,  
\end{eqnarray} 
the metric induced on the $x=0$ brane is
\begin{eqnarray}\label{eq:inducedmetric}
ds_b^2=\frac{1}{A^2}\left(-{f}(\rho)\,dt^2+{f}(\rho)^{-1} d\rho^2+\rho^2 d\phi^2\right)\, .
\end{eqnarray}
The interpretation of this geometry depends on whether $\mu$ vanishes or not. For  $\mu =0$, the function $f(\rho)$ is given by
\begin{eqnarray}
 {f}(\rho)=\lambda \rho^2-1 \,.\label{branestringmetric}
\end{eqnarray}
For $\lambda > 0$ and $\phi$ periodically identified with arbitrary period $\Delta \phi$, this corresponds to a `BTZ black string', analogous to that in \cite{Chamblin:1999by}, extending into the bulk. The full four-dimensional bulk metric for this string is obtained by setting $\mu = 0$ in (\ref{AdSCmetric}). Its event horizon is located at $\rho\equiv \rho_h=1/\sqrt{\lambda}$ and the metric function $G(x)$ in (\ref{AdSCmetric}) never vanishes. This black string corresponds to the dashed branch $(C)$ in the phase diagram of figure \ref{fig:phasediag}.

When $\mu \neq 0$, we can consider $\mu >0$ only without loss of generality. In this case, the metric (\ref{AdSCmetric}) describes a black hole with horizon located at $\rho\equiv\rho_h=-1/y_0 \equiv -1/y_h $, with $y_h \equiv y_0$ the negative root of $H(y)=0$. The region of interest  in this case is thus  further restricted to $y_0 \le y\le 0$. On the other hand, the function $G(x)$ has one, and only one, positive root $x_2$. This restricts the $x$ coordinate to range in the interval $0\le x \le x_2$. In order to avoid a conical singularity, the angle $\phi$ must now be given a period
\begin{eqnarray} \label{eq:period}
 \Delta\phi\equiv 2\pi \beta =4\pi/|G'(x_2)|. 
\end{eqnarray}
Finally, the black hole metric induced on the $x=0$ brane is (\ref{eq:inducedmetric}) with 
\begin{eqnarray}
f(\rho)=\lambda \rho^2-1-\frac{2\mu A}{\rho} \,.\label{branemetric} 
\end{eqnarray}
The metric  (\ref{eq:inducedmetric}), (\ref{branemetric}) induced on the brane looks like the BTZ black hole with $\mu A/\rho$ corrections. Indeed, the change of coordinates
\begin{eqnarray}\label{eq:rescaling}
t=\frac{A\,\Delta\phi}{2\pi}\hat{t}\,,\qquad \rho=\frac{2\pi A}{\Delta\phi}\hat{\rho}\,,\qquad \phi=\frac{\Delta\phi}{2\pi}\hat{\phi}\,,
\end{eqnarray}
exactly reproduces, in the $\mu \rightarrow 0$ limit, the canonical metric of the BTZ black hole \cite{Banados:1992wn} (see appendix \ref{app:BTZ}). Two further regimes can be distinguished in the $\mu >0$ case: small $\mu A$ (more precisely, $2\mu A<<\lambda^{-1/2}$) or large $\mu A$ ($2\mu A >> \lambda^{-1/2}$). The former case corresponds to a slightly corrected BTZ black hole on the brane, and is given by branch $(B)$ in the phase diagram of figure \ref{fig:phasediag}. Accordingly, for sufficiently small $\mu A$, branch $(B)$ approximates the string branch $(C)$. As $\mu A$ becomes larger, both branches start to differ. For $\mu A$ larger than a critical value, $M\sim (24 G_3)^{-1}$, branch $(B)$ disappears and a new branch $(A)$ arises. The latter corresponds to localized black holes heavily corrected from BTZ. Black holes on branch $(B)$ stick out significantly into the bulk, since they track the string branch $(C)$, while black holes in branch $(A)$ are flatter. See figure \ref{fig:phasediag} and table \ref{Table:parameters} for a summary.

Interestingly, the mass $M_3$ measured by a brane observer coincides with the four-dimensional mass \cite{Emparan:1999fd}
\begin{eqnarray} \label{eq:BHMass}
M \equiv M_3 = M_4 = \frac{1}{8G_3}\left(\frac{\Delta\phi}{2\pi}\right)^2 \; ,
\end{eqnarray}
when the relation between the four and three-dimensional Newton's constants, $G_3=A G_4 /2$, is taken into acount. Following convention, we usually set $G_3=1/8$ in the calculations of the next sections. The entropy and temperature for the BTZ black strings in branch $(C)$ are
\begin{eqnarray}
S_{bs}=\pi l_3\sqrt{\frac{2 M}{G_3}} \; , \qquad 
T_{bs}=\frac{\sqrt{2G_3 M}}{\pi l_3} \; , 
\end{eqnarray}
while for the black holes in branches $(A)$ and $(B)$ these are
\begin{eqnarray} \label{eq:SandT}
S_{bh}=\frac{2\pi}{ G_4 A^2\,|G'(x_2)|}\frac{x_2}{|y_0|(x_2+|y_0|)}  \; , \qquad 
T_{bh}=\frac{A}{2\pi}\frac{|H'(y_0)|}{|G'(x_2)|}  \; .
\end{eqnarray}
Recall that the period $\Delta \phi$ is unrestricted for the string, but is fixed to (\ref{eq:period}) for the black holes. Note also that the entropy and temperature (\ref{eq:SandT}) and the mass $M$ depend on the parameter $\mu$ through the root $x_2$ of the metric function $G(x)$ in (\ref{eqG}). This dependence leads to the two different branches $(A)$ and $(B)$ in the phase diagram of Figure \ref{fig:phasediag}. Besides these known, single center phases, other multicenter phases should exist composed of stable configurations containing a black hole localized on the brane and droplets in the bulk. We have denoted this phase as $(D)$ in figure \ref{fig:phasediag}. In principle, these multicenter configurations may exist for all values of the entropy and mass.

\begin{table*}\centering
\ra{1}
{\small
\begin{tabular}{@{}lcclccc@{}}\toprule
Parameter value or range    &  & & Description & & &  Branch 
\\ \midrule
$\mu = 0$    &  & & BTZ string & & &  (C) \\[10pt]
$2\mu A << \lambda^{-1/2}$    &  & & braneworld black hole sticking out into the bulk & & &  (B) \\[10pt]
$2\mu A >> \lambda^{-1/2}$    &  & & braneworld black hole well localized on the brane & & &  (A) \\[10pt]
all parameter ranges    &  & & multicenter configurations & & &  (D) \\
\bottomrule
\end{tabular}
}\normalsize
\caption{Black holes that arise for different values or ranges of the parameters in the line element (\ref{AdSCmetric}). The corresponding branches are depicted in figure \ref{fig:phasediag}.}
\label{Table:parameters}  
\end{table*}

Determining the stable phase in the diagram is therefore important, although we will not attempt to address this question in full generality. Instead, we will follow \cite{Fitzpatrick} to argue that certain multicenter configurations $(D)$, very close to each of the known branches $(A)$, $(B)$ and $(C)$, would provide stable endpoints of any potential instability, for example of Gregory-Laflamme type \cite{Gregory:1993vy,Gregory:2000gf}, that might affect the latter branches. The black string $(C)$ and the black hole $(B)$ might be more prone to instabilities than the flatter black hole branch $(A)$. However, even if these string and string-like branches were unstable to pinching off and decayed by emission of droplets into the bulk, the end state of the instability would not necessarily be branch $(A)$. The resulting multicenter, stable configurations $(D)$ are more likely to have an entopy very close to that of the black hole prior to the decay. Indeed, it was shown in \cite{Fitzpatrick} that the process of droplet emission is thermodynamically `marginal', in the sense that the entropy before and after the droplet formation is very similar. For large mass black holes, where the string $(C)$ is the only known phase, either this or a very close multicenter configuration $(D)$ would provide a stable phase. For small mass black holes, where the three phases $(A)$, $(B)$ and $(C)$ are known, this argument suggests the existence of stable, multicenter configurations very close to at least the latter two branches, which are already close to each other. Even if the more entropic and flatter branch $(A)$ is stable, it may coexist with these other multicenter configurations $(D)$ that are extremely close to $(B)$ and $(C)$.

We now move on to compute the quasi normal modes of black hole branches $(A)$ and $(B)$. Through appropriate coordinate rescalings, we can measure the quasinormal frequencies  in terms of the AdS$_3$ lengthscale.  Equivalently, we simply choose $l_3=1$, which in turn implies fixing $A=\lambda^{-1/2}$. We henceforth follow this convention.
  
\section{QNMs of brane-confined scalar perturbations}
\label{sec:QNMBrane}

In this section we compute the QNMs of the braneworld black holes we have just reviewed, associated to the perturbation with a scalar field confined to the $x=0$ brane. These would be the quasinormal frequencies measured by an observer living on the brane. This is, therefore, an entirely three-dimensional problem confined to the brane: the background geometry to be perturbed is the three-dimensional metric (\ref{eq:inducedmetric}) with (\ref{branemetric}), with the scalar perturbation taken to be of the form
\begin{eqnarray} \label{eq:scalarconfined}
 \Psi= e^{-i\omega t+im\phi \beta^{-1}}\,\rho^{-1/2}\,R(\rho)\,.
\end{eqnarray}
Here, $\omega$ is the frequency we want to determine, $m$ is an angular quantum number, the period $\beta$ was given in (\ref{eq:period}), and $R(\rho)$ is an arbitrary radial function. Although its formulation is entirely three-dimensional, our present problem does still know about the warped fourth dimension through the correction term $\mu A/\rho$ in the metric function $f(\rho)$ in (\ref{branemetric}). See \cite{Kanti:2005xa,Kanti:2006ua} for a similar setup in a related context. 

The wave equation,  $\nabla^2\Psi - M^2_\Psi  \Psi =0$, corresponding to a scalar field (\ref{eq:scalarconfined}) of mass $M^2_\Psi$ on the background (\ref{eq:inducedmetric}), (\ref{branemetric}) reduces to the Schr\"odinger-like radial equation
\begin{eqnarray}
\frac{d}{d\rho} \left[ f(\rho)  \frac{d}{d\rho}  R\right] +\left[\frac{\omega^2}{f(\rho)}  -\frac{M_\Psi^2}{A^2} -V(\rho)\right ] R  =0\,,\label{eq:BTZlike}
\end{eqnarray}
where we have defined the effective potential 
\begin{eqnarray} \label{poti2}
V(\rho)=\frac{\partial_{\rho}f(\rho)}{2\, \rho}-\frac{f(\rho)}{4\, \rho^2}+\frac{m^2}{\beta^2 \, \rho^2} \,.
\end{eqnarray}
Note that equation (\ref{eq:BTZlike}) depends on the parameter $\mu$ through $f(\rho)$ --see (\ref{branemetric}). For $\mu=0$, this QNM problem reduces to that for the BTZ black hole. QNMs for massless and massive scalar perturbations of BTZ have been computed in \cite{Cardoso:2001hn} and \cite{Birmingham:2001pj} respectively. Here, we are interested in the QNM problem for $\mu >0$, away from pure BTZ.

The QNM frequencies $\omega$ are obtained by solving the differential equation (\ref{eq:BTZlike}) with specific boundary conditions: outgoing at spatial infinity and ingoing at the horizon. In order to effectively implement these boundary conditions, we follow standard practice and introduce the tortoise coordinate
\begin{eqnarray}\label{eq:tortoise}
\rho_*=-\int\frac{d\rho}{\rho^2H(\rho)}=-\sum_{i=0}^2\frac{ \log|1-\rho_i/ \rho |}{H'(\rho_i)}\,.
\end{eqnarray}
We find it convenient to express this in terms of the function $H(\rho)$ (rather than $f(\rho)$ in  (\ref{branemetric})) obtained from (\ref{eqH}), (\ref{eq:rhocoord}). Here, $H'(\rho_i)$ is its derivative with respect to $\rho$, evaluated in the two zeroes of the latter: $H(\rho_i)=0$. In the tortoise coordinate, the horizon $\rho_h$ is mapped to $\rho_*\rightarrow -\infty$ and spatial infinity on the brane, $y=- \infty$, corresponds to $\rho_*\rightarrow0$. The boundary conditions are most easily imposed by further performing a change of function
\begin{align}\label{scalardef}
R(y) = e^{-i\omega\rho_*}\Phi(y) \; ,
\end{align}
so that ingoing modes at the event horizon, $\Psi(t,y,\phi) \sim e^{-i\omega(t+\rho_*)}$, and outgoing modes with (Dirichlet) boundary conditions at infinity, $\Psi(t,y,\phi) \sim e^{-i\omega(t-\rho_*)}$, respectively behave in the new function as 
\begin{eqnarray} \label{eq:BCsConfined}
\rho_*\rightarrow -\infty: \; \Phi(y) = O(1) \; , \quad \textrm{and} \quad 
\rho_*\rightarrow 0: \; \Phi(y) \sim e^{2i\omega \rho_*}\sim (y-y_h)^{2i\omega/H'(y_h)} \; .
\end{eqnarray}
 We present our results using Dirichlet boundary conditions at spatial infinity,  $\rho_*\rightarrow0$. We have nevertheless calculated the QNMs that arise from imposing Neumann boundary conditions. The results are qualitatively similar. Finally, since we will be interested in comparing these QNMs to those of the pure BTZ black hole, we rescale the coordinates $(t,\phi)$ as in (\ref{eq:rescaling}) and introduce
\begin{eqnarray}
\hat{\omega}=\omega \,A\,\beta \; , \qquad \hat{m}=m \; .
\end{eqnarray}
With these definitions, the radial equation (\ref{eq:BTZlike}) eventually becomes
\begin{align}\label{diffequation}
s(y)\,\Phi''(y) + t(x) \,\Phi'(y) +u(y)\, \Phi(y)=0\,,
\end{align}
where
\begin{align}
s(y) \equiv H(y)\,,\quad
t(y) \equiv H'(y) + \frac{2i\hat{\omega}}{A\beta}\,,\quad
u(y) \equiv \frac{\hat{m}^2}{\beta^2} - \frac{M^2_\Psi}{A^2 y^2}+\frac{2 y H'(y)-3 H(y)}{4 y^2},
\end{align}
and we have found it convenient to express this in terms of the function $H(y)$ in (\ref{eqH}).

\begin{figure}
\centering
\begin{minipage}[t]{.4\textwidth}
\centering
\vspace{0pt}
\includegraphics[width=\textwidth]{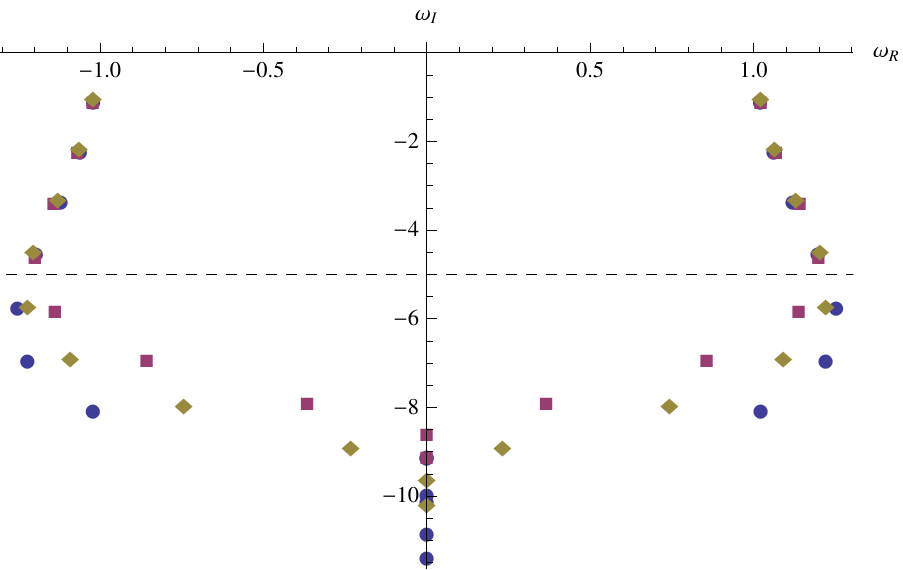}
\end{minipage}
\hspace{1cm}
\begin{minipage}[t]{.4\textwidth}
\centering
\vspace{0pt}
\begin{tabular}{c c c } 
\hline\hline $n$ &$\hat{\omega}_{R \, n}$ & $\hat{\omega}_{I \, n}$ \\ [0.5ex]	
\hline 
1&$\pm 1.021$&$ -1.117$ \\ 
2&$\pm 1.062$&$-2.247$ \\
 3&$\pm 1.120$&$-3.387$  \\
 4 & $\pm 1.196 $&$ -4.552$  \\ 
 [1ex] \hline 
\end{tabular}
\end{minipage}
\caption{QNMs $\hat{\omega}_{I \, n}$ vs $\hat{\omega}_{R \, n}$ for small $\mu A=0.65$ (on the left), $\hat{m}=0$ and fixed $\lambda=0.001$ ($A=1/\sqrt{\lambda}$), with increasing overtones depicted downwards. The convergence of the numerical method ($N=16$ in red, $N=18$ in yellow, $N=20$ in blue) is also represented: the modes above the dotted line are the trustworthy (up to three decimal digits) and given explicitly in the table on the right.} \label{Fig:RealvsImBulk}
\end{figure}

We now solve the boundary value problem (\ref{diffequation}), (\ref{eq:BCsConfined}) using the Frobenius method. Namely, we expand the coefficients of the differential equation (\ref{diffequation}) and its two linearly-independent solutions $\Phi^{(i)}(y)$, $i=1,2$, in powers  of $y$ around the horizon $y=y_h$,
\begin{eqnarray} \label{eq:expansionConfined}
\Phi^{(i)}(y) = (y-y_h)^{\gamma_{i}} \sum_n a_n (y-y_h)^n \; , \qquad 
s(x) = \sum_{n=0}^N s_n (x-x_2)^n \; , 
\end{eqnarray}
and similarly for $t(x)$ and $u(x)$. Here, $N$ is a sufficiently large integer (that is, infinite up to numerical precision), and the exponents $\gamma_i$,
\begin{eqnarray} \label{eq:IndicialExponents}
\gamma_1 =0 \; , \qquad 
\gamma_2 =  \frac{2i\hat{\omega}}{H'(y_h)} \; , 
\end{eqnarray}
are the solutions to the indicial equation
\begin{eqnarray}
\gamma(\gamma-1) +\gamma p_0+ q_0=0 \; ,  \quad 
p_0 \equiv \lim_{x\rightarrow x_2} \, (x-x_2)\frac{ t(x)}{s(x)} \; ,  \quad 
q_0 \equiv \lim_{x\rightarrow x_2} \, (x-x_2)^2\frac{ u(x)}{s(x)} \; .
\end{eqnarray}
Substituting into the differential equation (\ref{diffequation}), the following exact recursion relation for the unknown coefficients $a_k$ is found:
\begin{eqnarray} \label{eq:CoeffsConfined}
a_n = -\frac{1}{P_n} \sum_{k=0}^{n-1} \left [s_{n-k}k(k-1) + t_{n-k}k + u_{n-k}\right ]\, a_k\,,
\end{eqnarray}
where $a_0$ can be set to one without loss of generality (as long as the eigenfunction's normalization is left arbitrary), and
\begin{eqnarray} \label{eq:DefP}
P_n = n(n-1)s_0 + n t_0 + u_0 \; .
\end{eqnarray}

The functions $\Phi^{(i)}(y)$ in (\ref{eq:expansionConfined}) with the coefficients (\ref{eq:CoeffsConfined}) solve the differential equation (\ref{diffequation}). We next impose the two boundary conditions (\ref{eq:BCsConfined}). The exponents $\gamma_1$ and $\gamma_2$ respectively lead to ingoing and outgoing waves at the black hole horizon. We thus discard the solution $\Phi^{(2)}(y)$ and keep the solution $\Phi^{(1)}(y)$, with $\gamma_1 = 0$. Imposing the second boundary condition, at spatial infinity, finally leads to a discretization of the possible values
\begin{eqnarray} \label{eq:QNMsConfined}
\hat \omega_n \equiv  \hat \omega_{R \, n} + i \,  \hat \omega_{I \, n}  \; , \quad  n=1,2, \ldots
\end{eqnarray}
of the frequencies compatible with this boundary condition. These are the QNMs of the braneworld black hole (\ref{eq:inducedmetric}), (\ref{branemetric}) subject to the brane-confined perturbation (\ref{eq:scalarconfined}). Typical values of these QNMs and a plot in the complex plane are given in figure \ref{Fig:RealvsImBulk}.

\begin{figure}
    \centering
    \includegraphics[width=0.45\textwidth]{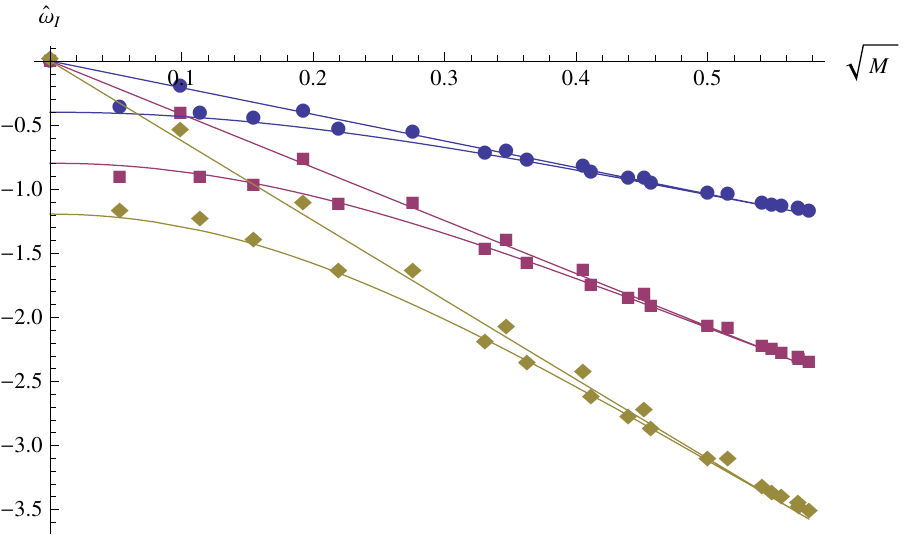}
    \caption{The first three overtones $(n=1,2,3=blue,red,yellow)$ of the imaginary part of the QNMs $\hat{\omega}_{I \, n}$ corresponding to the black holes (\ref{eq:inducedmetric}), (\ref{branemetric}) on the brane, perturbed by a massless scalar field (\ref{eq:scalarconfined}) confined to  the brane. The QNMs are shown as a function of the square root of the black hole mass $M$, at fixed $\hat{m}=1$ and $\lambda=0.001$ ($A=1/\sqrt{\lambda}$). The interpolating lines correspond to fits proportional to the entropy according to equation (\ref{eq:QNMsEntropy}). The upper and lower branches correspond to small and large $\mu A$ (branches $(B)$ and $(A)$ in the phase diagram Fig. \ref{fig:phasediag}).}\label{BTZlike}
\end{figure}

In figure \ref{BTZlike} we plot the first three overtones, $n=1,2,3$, of the imaginary part $ \hat \omega_{I \, n}$ of the QNMs (\ref{eq:QNMsConfined}) for vanishing scalar mass $M_\Psi^2 = 0$, fixed values of the cosmological constant on the brane and of the angular quantum number $\hat m$, and various values of the square root of the black hole mass $M$, (\ref{eq:BHMass}), (\ref{eq:period}). Remarkably enough, this plot reflects a common pattern for the QNMs of both black hole branches $(A)$ and $(B)$. Upon appropriate rescaling and overall sign change, the plot \ref{BTZlike} can be brought to coincide, up to numerical precision, with the black hole branches of the phase diagram in figure \ref{fig:phasediag}. In other words, the QNMs of both branches $(A)$ and $(B)$ go linearly with the overtone and with the black hole entropy (\ref{eq:SandT}). We numerically find that
\begin{eqnarray} \label{eq:QNMsEntropy}
\hat{\omega}_{I \, n} \approx - \frac{ n+1 }{ 6.04\pm0.02 } \, S_{bh} \; 
\end{eqnarray}
fits well the data. If the fit is restricted to QNMs for small $\mu A$ in branch $(B)$, the proportionality constant approaches $1/(2\pi)$, like for the BTZ black hole.

See section \ref{sec:Discussion} for further discussion.

\section{QNMs of bulk-probing scalar perturbations}
\label{sec:QNM}

Determining the stable black hole configuration among all possible phases is an interesting question in our context. In this section, we check for potential instabilities of the braneworld black holes against perturbations that are also allowed to probe the warped fourth dimension. We again focus on scalar perturbations for simplicity. For these perturbations, we rather find stability as well, and compute the corresponding QNMs. Of course, there are other types of perturbations that may lead to instabilities, so it would be interesting to extend the present analysis to include perturbations of other spins.

The natural starting point for this analysis is the full four-dimensional geometry (\ref{AdSCmetric}) equipped with the brane construction \cite{Emparan:1999fd} reviewed in section \ref{sec:Setup}. As discussed in  \cite{Nozawa:2008wf}, the wave equation on the background  (\ref{AdSCmetric}) for a scalar that is conformally coupled to gravity becomes separable\footnote{In  \cite{Nozawa:2008wf}, QNMs are calculated for the $k=-1$ black holes of  \cite{Emparan:1999fd} . Recall that here we are interested in the $k=1$ case.}. We will thus analyze the QNMs associated to the wave equation
\begin{eqnarray}\label{eom}
\nabla^2 \Psi -\frac{1}{6}\mathcal{R} \Psi=0\,,
\end{eqnarray} 
where $\mathcal{R}=-12 A^2 (1+\lambda)$  is the Ricci scalar of the metric (\ref{AdSCmetric}), with $\lambda >0$. The wave equation (\ref{eom}) is conformally invariant, and becomes separable after performing a conformal transformation on $\Psi$ and the background metric  (\ref{AdSCmetric}) that essentially removes its overall factor $A^2 (x-y)^2$: see  \cite{Nozawa:2008wf} for the details. Substituting the ansatz
\begin{eqnarray} \label{eq:ScalarBulk}
 \Psi= e^{-i\omega t+im\phi \beta^{-1}}\,R(y)\,T(x)\,,
\end{eqnarray}
into the conformally-transformed equation, we obtain the following separate equations for the functions $T(x)$ and $R(y)$,
\begin{eqnarray}
&&\frac{d}{dx} \left[ G(x) \frac{d}{dx} T \right] + \left[ K-\frac{m^2 }{\beta^{2}G(x)} -2 \mu A \,x \right ] T=0\,,\label{eq:sheroidal}\\
&&\frac{d}{dy} \left[ H(y)  \frac{d}{dy}  R\right] +\left[ K+\frac{\omega^2}{H(y)}  -2 \mu A \,y \right ] R =0\, ,\label{eq:radial}
\end{eqnarray}
where $K$ is the separation constant. We will respectively refer to (\ref{eq:sheroidal}) and (\ref{eq:radial}) as the spheroidal, or angular, and radial equations.

Both ODEs (\ref{eq:sheroidal}) and (\ref{eq:radial}) have four regular singular points at $x=\{\infty,x_i\}$ for $i=0,1,2$ where $G(x_i)=0$, and at $y=\{\infty,y_i\}$ for $i=0,1,2$ where $H(y_i)=0$.  For convenience, we will accordingly rewrite the metric functions (\ref{eqH}), (\ref{eqG}) as  
\begin{eqnarray}
G(x)=(-2\mu A)(x-x_0)(x-x_1)(x-x_2) \; , \quad 
H(y)=(-2\mu A)(y-y_0)(y-y_1)(y-y_2)  .
\end{eqnarray}
Note, in particular, that the root $y_0 \equiv y_h$ is real, while $y_1$, $y_2$ are complex. We first determine the eigenvalues $K$ by giving suitable boundary conditions to the spheroidal equation, and then insert these values in the radial equation in order to solve for the quasinormal frequencies. We consider both equations in turn.

\subsection{Angular Equation}

Following \cite{Nozawa:2008wf}, we impose boundary conditions for $T(x)$ such that the function is regular at $x = x_2$, where $x_2$ is the single positive root of $G(x)$, and Neumann boundary conditions $T'(0) = 0$ at the position of the brane, $x=0$. This boundary condition reflects a $\mathbb{Z}_2$ symmetry about the brane. We can obtain numerical values for the eigenvalue $K$ of equation (\ref{eq:sheroidal}) by expanding the solutions $T(x)$ in a power series about $x=x_2$, and then enforcing the $\mathbb{Z}_2$ symmetry boundary condition (i.e. the Neumann boundary condition) at $x=0$, term by term in the power series, so that each additional term gives greater accuracy. To carry this out, we proceed as in section \ref{sec:QNMBrane}. We thus define  $T(x)=\Phi(x)$ and write the spheroidal equation (\ref{eq:sheroidal}) as in (\ref{diffequation}), where now 
\begin{align}
s(x) \equiv G(x)^2\,,\quad
t(x) \equiv G'(x)G(x)\,, \quad
u(x) \equiv G(x)(K - 2\mu Ax)-\frac{m^2}{\beta^2}\, . 
\end{align}
We expand these functions in powers of $x$ around $x=x_2$, so that $s(x) = \sum_{n=0}^N s_n (x-x_2)^n$, etc., and determine the indicial exponents to be $\gamma_{1,2}=\pm m/2$. Since, in this case, $\gamma_1$, $\gamma_2$ differ by an integer, Frobenius' method asserts that 
the two linearly independent solutions to (\ref{diffequation}) can be written as 
\begin{align}
T^{(1)}(x) &=  (x-x_2)^{\gamma_1}\sum_{n}b_n^{(1)}(x-x_2)^n\ \label{eq:solution1}\,,\\
T^{(2)}(x) &= T^{(1)}(x)  \log (x-x_2) + (x-x_2)^{\gamma_2}\sum_{n}b_n^{(2)}(x-x_2)^n\,.
\end{align}

The general solution to (\ref{diffequation}) is thus the linear combination $T(x)=c_1 T^{(1)}(x)+ c_2 T^{(2)}(x)$ for arbitrary constants $c_1,c_2$. In order to ensure regularity of $T(x)$ at $x=x_2$, we set $c_2 =0$ and keep $c_1 \neq 0$. Inserting (\ref{eq:solution1}) into the differential equation, we find the recursion relation (\ref{eq:CoeffsConfined}), (\ref{eq:DefP}), with the $a$'s replaced by the $b$'s. We then solve numerically for $K$ by imposing the $\mathbb{Z}_2$ symmetric boundary condition specified above. We do this by truncating the infinite sum (\ref{eq:solution1}) to a sufficiently large number of terms $N$, for fixed angular quantum number $m$. As we take $N$ larger and larger, we get a better approximation to the eigenfunction $T(x)$ and eigenvalue $K$. Note, incidentally, that the series (\ref{eq:solution1}) converges in the range $0\le x\le x_2$ since the latter  belongs to the region within the radius of convergence  $x<|x_2-x_1|$ or $x<|x_2-x_0|$. The first few angular eigenvalues we find, for parameters fixed {\it e.g.} to $\mu A=3$ and $m=2$, are
\begin{eqnarray} \label{eq:AngularEigenvalue}
K= \nu(\nu+1) \; , \quad \textrm{with} \quad 
\nu=\{5.658, \; 8.770, \; 12.346, \; 16.316 , \;  \ldots \} \; .
\end{eqnarray}
%

 \subsection{Radial Equation}

Having determined the eigenvalues $K$ of the angular equation, we now insert these into the radial equation (\ref{eq:radial}) in order to determine, for each $K$, the quasinormal frequencies $\omega$ by solving the associated boundary value problem. 

In terms of the tortoise coordinate $\rho_*$ defined in (\ref{eq:tortoise}), the radial ODE (\ref{eq:radial}) can be rewritten as
\begin{eqnarray} \label{Eqrad}
\left[\frac{d^2}{d\rho_*^2}+ \omega^2 - V(\rho) \right] R=0\,,
\end{eqnarray}
where the effective potential is now
\begin{eqnarray} \label{poti}
V(\rho)= \Big( \lambda -\frac{1}{\rho^2} - \frac{2\mu A}{\rho^3}  \Big)  \Big( K+\frac{2 \mu A}{\rho}  \Big) .
\end{eqnarray}
This resembles the potential that arises in the radial equation for minimally coupled scalar fields in a Schwarzschild-de Sitter background, see appendix \ref{app:dSSch}. The potential (\ref{poti})  vanishes solely at the event horizon $\rho_h=-1/y_h$, where  $\lambda\, \rho_{h}^3-\rho_{h} -2\mu A=0$. 

In order to impose the boundary conditions it is natural to proceed as in section \ref{sec:QNMBrane} and define the new radial function (\ref{scalardef}). The radial equation (\ref{eq:radial}) then becomes (\ref{diffequation}), now with
\begin{align} \label{eq:CoeffsBulk}
s(y) \equiv H(y)\,,\qquad
t(y) \equiv H'(y) + \frac{2i\hat{\omega}}{A\beta}\,,\qquad
u(y) \equiv K - 2\mu Ay \; .
\end{align}
We again make use of the Frobenius method in order to solve for the quasinormal modes.  We thus expand the two linearly-independent solutions $\Phi^{(i)}(y)$, $i=1,2$, to the radial ODE and the coefficient functions (\ref{eq:CoeffsBulk}) around the horizon $y=y_h$, as in (\ref{eq:expansionConfined}), with indicial exponents (\ref{eq:IndicialExponents}). Note that the radius of convergence of these series contains the physically sensible range $y_h \leq y \leq 0 $. We discard the solution $\Phi^{(2)}(y)$ and retain only $\Phi^{(1)}(y)$, with $\gamma_1 = 0$, in order to obtain a purely ingoing wave at the horizon. The recursion relation satisfied by the coefficients of $\Phi^{(1)}(y)$ in the expansion (\ref{eq:expansionConfined}) is again (\ref{eq:CoeffsConfined}) with (\ref{eq:DefP}). Notice that, in this case, the recursion relation depends on the angular eigenvalue $K$ through the expanded functions (\ref{eq:CoeffsBulk}). Taking for $K$ the values (\ref{eq:AngularEigenvalue}) obtained at fixed parameters, the only quantity that remains to be specified is the frequency $\hat \omega$. This is finally fixed by imposing (Dirichlet) boundary conditions at spatial infinity, $y=0$. 

\begin{figure} [t]
    \centering
    \includegraphics[width=0.45\textwidth]{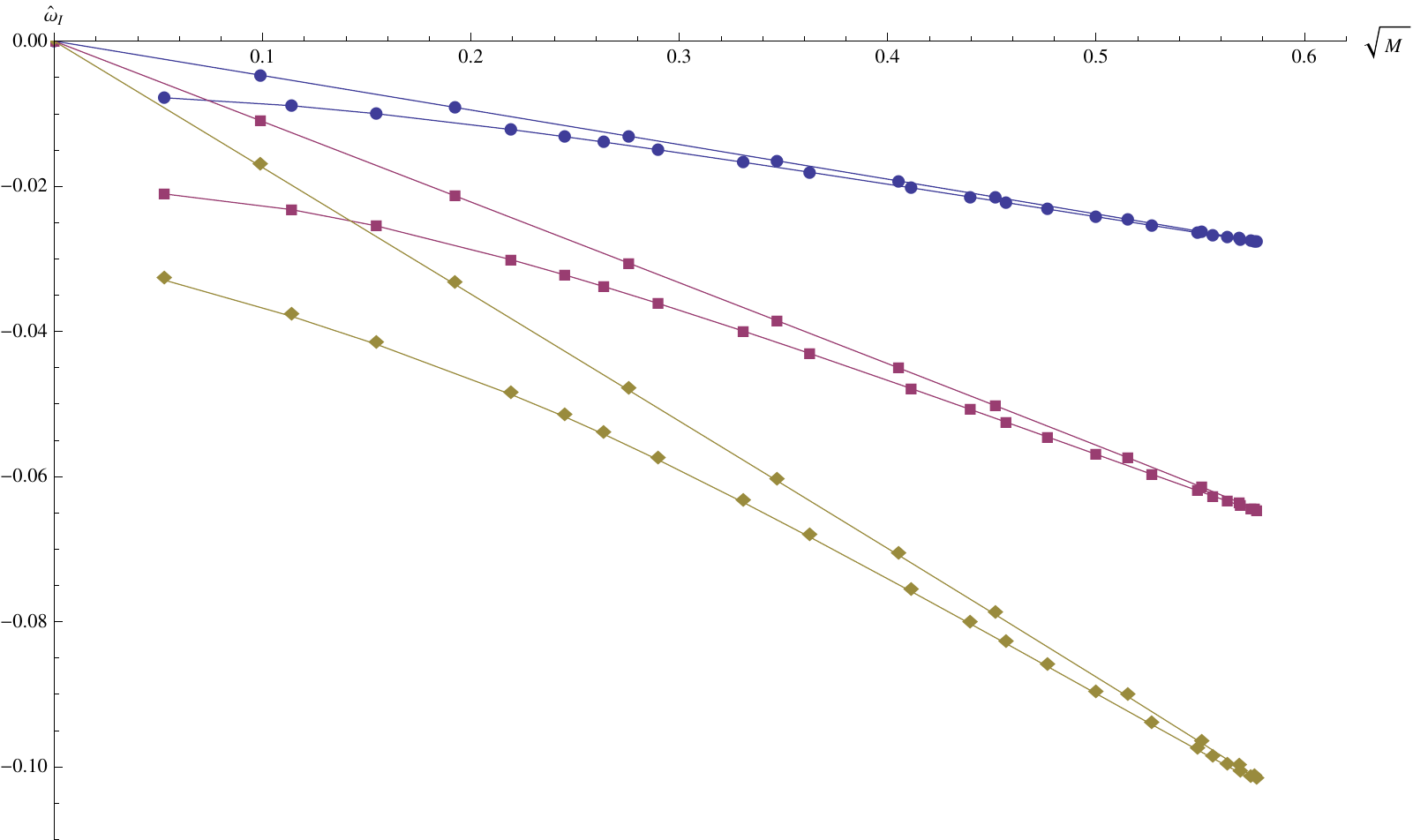}
    \caption{The first overtones $(n=1,2,3=blue,red,yellow)$ of the imaginary parts of the quasinormal modes of bulk-probing perturbations (\ref{eq:ScalarBulk}) as a function of the black hole mass $\sqrt{M}$ - (right) $\hat{\omega}_{R \, n} $ vs $\sqrt{M}$ and (left) $\hat{\omega}_{I \, n}$ vs $\sqrt{M}$ for different values of $\mu A$ and fixed (lowest lying) eigenvalue $K$, $\hat{m}=1$, $\lambda=0.001$ ($A=1/\sqrt{\lambda}$). The upper and lower branches correspond to small and large $\mu A$ (branches $(B)$ and $(A)$ in the phase diagram Fig. \ref{fig:phasediag}).} \label{fig:QNMs3}
\end{figure}

The frequencies compatible with these boundary conditions again correspond to a discrete set of the form (\ref{eq:QNMsConfined}). These are the  QNMs of scalar perturbations (\ref{eq:ScalarBulk}) conformally-coupled to the braneworld black holes obtained from (\ref{AdSCmetric}). We have written the first few overtones in table \ref{table1} and have plotted their imaginary parts against the square root of the black hole mass in figure \ref{fig:QNMs3}. Also for these perturbations, the imaginary part of the QNMs tracks down the entropy of the corresponding phase, with a proportionality constant that differs from (\ref{eq:QNMsEntropy}).

\begin{table}[!htbp] \centering 
\begin{tabular}{c c c } 
\hline\hline $n$ &$\hat{\omega}_{R \, n}$ & $\hat{\omega}_{I \, n}$ \\ [0.5ex]	
\hline 
1&$\pm 0.0860$&$ -0.0589$ \\ 
2&$\pm 0.1121$&$-0.1471$ \\
 3&$\pm 0.1425$&$-0.2371$  \\
 4 & $\pm 0.1744 $&$ -0.3265$  \\ 
 [1ex] \hline 
\end{tabular}
\caption{QNMs of (conformally coupled) massless scalar field for the lowest $K$ eigenvalues, $\hat{m}=0$, fixed $\lambda=0.001$ and $\mu A=25$.} 
 \label{table1} 
\end{table}

\section{Discussion and comparison to BTZ}
\label{sec:Discussion}

In this paper we have computed QNMs of static black holes on negatively curved three-dimensional branes. The oscillation caused by the perturbations we consider are stable and damped. Indeed, the discrete set of quasinormal frequencies that we find have a real part, describing the oscillation, and a negative imaginary part, describing its ring down. We find a common pattern for the QNMs of all known branches of braneworld black holes as calculated in the classical limit: their imaginary parts are proportional to the overtone and to the corresponding black hole entropy. For the BTZ black hole, for which the scalar QNMs are known \cite{Cardoso:2001hn}, this proportionality also holds: see equations (\ref{eq:QNMsBTZ}), (\ref{eq:TandSBTZ}). This proportionality thus extends to all other black holes subject to the purely three-dimensional (that is, brane-confined) scalar perturbations that we considered in section \ref{sec:QNMBrane}.

An asymptotic observer in three-dimensional AdS space that sent waves to a black hole and determined the corresponding QNMs, would thus obtain a measure of the black hole entropy. And which entropy she measured would depend on which is the stable black hole phase. For `astrophysical' black holes of large mass, the BTZ black string $(C)$ is the only known solution in the phase diagram. If this solution is stable, an observer would exactly measure the QNMs of the BTZ black hole, to which the string reduces on the brane. Even if branch $(C)$ were unstable\footnote{Instabilities have been argued to exist in \cite{Liu:2008ds}.} and decayed  by droplet emission into the bulk, the entropy of the resulting multicenter configuration $(D)$ would still be very similar to that of the black string $(C)$, as we argued in section \ref{sec:Setup}. Under the reasonable assumption that the quasinormal modes of this multicenter configuration also follow the entropy, an observer would thus still measure QNMs that are experimentally indistinguishable from those of the BTZ black hole. We thus reach the interesting conclusion that fully non-linear gravitational effects with and without brane are indistinguishable for large black holes, as far as this particular experiment is concerned.

The situation is of course more subtle for small black holes with mass of the order of the three-dimensional Planck mass. In this case, there do exist two branches of black holes, $(A)$ and $(B)$, in addition to the BTZ black string $(C)$. Being proportional to the entropy, the QNMs for black holes in branches $(A)$, flatter and heavily corrected from BTZ, and $(B)$, stickier into the bulk and closer to BTZ, considerably differ. If an observer measured QNMs corresponding to a black hole in branch $(A)$ rather than those of the braneless BTZ black hole, she could immediately conclude the existence of the brane. However, recall from section \ref{sec:Setup} our argument that branch $(A)$ might not be the only stable configuration or might not be stable at all. Other stable, multicenter black holes $(D)$, very similar to BTZ and well localized on the brane might exist. These would dominate the endpoint of the presumable instability of branch $(B)$ to decay by droplet emission.  For sufficiently small mass, where branches $(B)$ and $(C)$ are very close, a conclusion similar to that for the `astrophysical' black holes could be reached: also for black holes in this branch would the full non-linear effects of gravity on this braneworld experiment be insensitive to the presence of the brane.

Our test for instabilities in section \ref{sec:QNM} was inconclusive, although we did not intend it to be exhaustive. Further work to determine the stable phase space could therefore be very interesting. More generally, it is of interest to learn more about nonlinear solutions in RSII type braneworlds to test how closely physical quantities resemble the lower-dimensional world.

\subsection*{Acknowledgements}

We thank Geoffrey Comp\`ere, Gary Horowitz, Alex Maloney, Andy Strominger and Jennie Traschen for interesting discussions. This work was supported in part by the Fundamental Laws Initiative at Harvard. LR is supported in part by NSF grants PHY-0855591 and PHY-1216270. OV is supported by the Marie Curie fellowship PIOF-GA-2012-328798, managed from the CPHT of \'Ecole Polytechnique. MJR and OV would also like to thank the Centre de Recherches Math\'ematiques, Montreal, for hospitality during the workshop {\it AdS/CFT, self-adjoint extension and the resolution of cosmological singularities}.

\appendix

\section{QNMs for the BTZ Black Hole}
 \label{app:BTZ}
 
The QNMs modes of the BTZ black hole subject to massless, minimally coupled scalar perturbations have been computed in \cite{Cardoso:2001hn}. Here, we easily rederive their results using the monodromy method developed in \cite{Castro:2013kea}.

The BTZ black hole \cite{Banados:1992wn} is a solution to three-dimensional Einstein gravity with a negative cosmological constant,
 \begin{eqnarray}
 R_{\mu\nu} = -  2\, l_3^{-2} g_{\mu \nu} \, .
 \end{eqnarray}
 The line element of the static solution is 
 \begin{eqnarray}
ds_{BTZ}=-f(\hat{\rho})d\hat{t}^2+f(\hat{\rho})^{-1} d\hat{\rho}^2+\hat{\rho}^2 d\hat{\phi}^2\,,\qquad \text{where} \quad f(\hat{\rho})= \frac{\hat{\rho}^2}{l_3^2} -M_3\, ,
\end{eqnarray}
and its temperature and entropy are given in terms of its mass $M_3$ by
 \begin{eqnarray} \label{eq:TandSBTZ}
T_{BTZ}=-\frac{\sqrt{M_3}}{2\pi l_3} \,,\qquad S_{BTZ}=4\pi l_3\sqrt{ M_3} \; .
\end{eqnarray}

In order to study the QNMs for massless, minimally coupled scalar perturbations to the BTZ black hole,
\begin{eqnarray} \label{eq:KGBTZ}
\nabla^2 \Psi =0\, ,
\end{eqnarray}
we have to impose that the transmission coefficient vanishes, $|\mathcal{T}|^{-2}=0$, as required by the usual QNM boundary conditions. This in turn implies
\begin{eqnarray}
\alpha_3+\alpha_1-\alpha_2=0,\qquad \alpha_3-\alpha_1+\alpha_2=0 \; , 
\end{eqnarray}
where
\begin{eqnarray}
\{\alpha_1,\alpha_2,\alpha_3\}=\{\frac{i\omega}{\sqrt{M_3}},2, \frac{im}{\sqrt{M_3}}\}
\end{eqnarray}
are the monodromies at the regular singular points $0,1,\infty$ of the radial differential equation that arises from (\ref{eq:KGBTZ}). We thus immediately obtain
\begin{eqnarray} \label{eq:QNMsBTZ}
\omega l_3=\pm m-2i(n+1) M_3^{1/2}/l^2_3
\end{eqnarray}
where $m$ is the angular quantum number and $n$ the tone number, in agreement with \cite{Cardoso:2001hn}.

\section{Schwarzschild-de Sitter radial potential}
 \label{app:dSSch}

For the bulk-probing perturbations of section \ref{sec:QNM}, the QNMs we obtain resemble qualitatively the QNMs for Schwarzschild-de Sitterblack holes \cite{Zhidenko:2003wq}. This is likely due to the fact that the radial wave equations of both problems can be mapped to one another by a redefinition of the physical parameters that enter the equations. A similar observation was made in \cite{Nozawa:2008wf} about the QNMs of other class of braneworld black holes. 

It is thus interesting to compare the radial potential (\ref{poti}) that arises for the QNM problem of section \ref{sec:QNM} with that obtained for the analogue QNM problem on a four-dimensional Schwarzschild-(Anti)-de Sitter black hole,
\begin{eqnarray} \label{SchdS}
ds^2 = -f(\rho) dt^2 + f(\rho)^{-1} d\rho^2 +\rho^2 (d\theta^2 + \sin^2 \theta d\phi^2) \, , \qquad f(\rho)  = 1-\frac{2M}{\rho} -\frac{\Lambda}{3} \rho^2 , 
\end{eqnarray}
of mass $M$ and cosmological constant $\Lambda$. As shown in {\it e.g.}~\cite{Crispino:2013pya}, the radial equation corresponding to a scalar conformally coupled  ($\xi = \frac{1}{6}$  in  \cite{Crispino:2013pya}) to the black hole (\ref{SchdS}) is again (\ref{Eqrad}) but now with
\begin{eqnarray}
V(\rho)= \Big( -\frac{\Lambda}{3} +\frac{1}{\rho^2} - \frac{2M}{\rho^3}  \Big)  \Big( \tilde{K}+\frac{2M}{\rho}  \Big) .
\end{eqnarray}
We thus see that the potential (\ref{poti}) coincides with the potential of a conformally coupled scalar in a Schwarzschild-de Sitter black hole of mass $M = -\mu A$, separation constant $\tilde{K}=-K$ and cosmological constant $\Lambda = 3\lambda >0$. For $\mu A >0$, the potential (\ref{poti}) is thus mapped to that of an unphysical, negative mass Schwarzschild-de Sitter black hole. This is unlike  \cite{Nozawa:2008wf}, where the radial equation of the brane-world black-hole considered therein does reduce to that of (asymptotically flat) Schwarzschild. 



\begin{thebibliography}{99}

  
  
\bibitem{RandallSundrumI}
L.~Randall and R.~Sundrum, {\em {Large Mass Hierarchy from a Small Extra Dimension}},
\newblock Phys. Rev. Lett. {\bf 83}, 3370, 1999 [arXiv:hep-ph/9905221].

\bibitem{RandallSundrumII}
L.~Randall and R.~Sundrum, {\em {An Alternative to Compactification}}, 
\newblock Phys. Rev. Lett. {\bf 83}, 4690, 1999 [arXiv:hep-th/9906064].



\bibitem{RSLinGrav1} 
S.~B.~Giddings, E.~Katz and L.~Randall, {\em {Linearized gravity in brane backgrounds}},
\newblock JHEP {\bf 0003}, 023, 2000 [arXiv:hep-th/0002091]

\bibitem{RSLinGrav2} 
S.~B.~Giddings and E.~Katz, {\em {Effective theories and black hole production in warped compactifications}},
\newblock J. Math. Phys. {\bf 42}, 3082, 2001 [arXiv:hep-th/0009176].

\bibitem{RSLinGrav3} 
N.~Arkani-Hamed, M.~Porrati and L.~Randall, {\em {Holography and phenomenology}},
\newblock JHEP {\bf 08}, 17, 2001 [arXiv:hep-th/0012148].

\bibitem{Gregory:2008rf}
  R.~Gregory,
  {\em Braneworld black holes},
  Lect.\ Notes Phys.\  {\bf 769} (2009) 259
  [arXiv:0804.2595 [hep-th]].



\bibitem{Tanahashi:2011xx}
  N.~Tanahashi and T.~Tanaka,
  {\em Black holes in braneworld models},
  Prog.\ Theor.\ Phys.\ Suppl.\  {\bf 189} (2011) 227
  [arXiv:1105.2997 [hep-th]].
   


\bibitem{Tanaka} 
T.~Tanaka, {\em {Classical black hole evaporation in Randall-Sundrum infinite braneworld}},
\newblock Prog. Theor. Phys. Suppl. {\bf 148}, 307 (2003) [arXiv:gr-qc/0203082].

\bibitem{EmparanConj} 
R.~Emparan, A.~Fabbri, and N.~Kaloper, {\em {Quantum black holes as holograms in $AdS$ braneworlds}},
\newblock JHEP {\bf 0208}, 043 (2002) [arXiv:hep-th/0206155].

\bibitem{Fitzpatrick} 
A.L.~Fitzpatrick, L.~Randall, and T.~Wiseman, {\em {On the existence and dynamics of braneworld black holes}},
\newblock JHEP {\bf 0611}, 033, 2006 [arXiv:hep-th/0206155].

\bibitem{Wiseman} 
P.~Figueras and T.~Wiseman, {\em {Gravity and large black holes in Randall-Sundrum II braneworlds}},
\newblock Phys. Rev. Lett. {\bf 107}, 081101, 2011 [arXiv:1105.2558].

\bibitem{Page} 
S.~Abdolrahimi, C.~Catto\"{e}n, D.~N.~Page, S.~Yaghoobpour-Tari, {\em {Spectral Methods in General Relativity and Large Randall-Sundrum II Black Holes}},
\newblock JCAP {\bf 1306}, 039, 2013 [arXiv:1212.5623].


  
\bibitem{Emparan:1999wa} 
  R.~Emparan, G.~T.~Horowitz and R.~C.~Myers,
 {\em Exact description of black holes on branes},
  JHEP {\bf 0001}, 007 (2000)
  [hep-th/9911043].



\bibitem{Emparan:1999fd} 
  R.~Emparan, G.~T.~Horowitz and R.~C.~Myers,
  {\em Exact description of black holes on branes. 2. Comparison with BTZ black holes and black strings},
  JHEP {\bf 0001}, 021 (2000)
  [hep-th/9912135].


\bibitem{Banados:1992wn}
  M.~Banados, C.~Teitelboim and J.~Zanelli,
  {\em The Black hole in three-dimensional space-time},
  Phys.\ Rev.\ Lett.\  {\bf 69} (1992) 1849
  [hep-th/9204099].

\bibitem{Hartnoll:2008vx}
  S.~A.~Hartnoll, C.~P.~Herzog and G.~T.~Horowitz,
  {\em Building a Holographic Superconductor},
  Phys.\ Rev.\ Lett.\  {\bf 101} (2008) 031601
  [arXiv:0803.3295 [hep-th]].

\bibitem{Cardoso:2001hn} 
  V.~Cardoso and J.~P.~S.~Lemos,
  {\em Scalar, electromagnetic and Weyl perturbations of BTZ black holes: Quasinormal modes},
  Phys.\ Rev.\ D {\bf 63}, 124015 (2001)
  [gr-qc/0101052].


\bibitem{Birmingham:2001pj} 
  D.~Birmingham, I.~Sachs and S.~N.~Solodukhin,
  {\em Conformal field theory interpretation of black hole quasinormal modes},
  Phys.\ Rev.\ Lett.\  {\bf 88}, 151301 (2002)
  [hep-th/0112055].


\bibitem{Chan:1996yk} 
  J.~S.~F.~Chan and R.~B.~Mann,
  {\em Scalar wave falloff in asymptotically anti-de Sitter backgrounds},
  Phys.\ Rev.\ D {\bf 55}, 7546 (1997)
  [gr-qc/9612026].
  

\bibitem{Kanti:2005xa}
  P.~Kanti and R.~A.~Konoplya,
  {\em Quasi-normal modes of brane-localised standard model fields},
  Phys.\ Rev.\ D {\bf 73} (2006) 044002
  [hep-th/0512257].
  
\bibitem{Kanti:2006ua}
  P.~Kanti, R.~A.~Konoplya and A.~Zhidenko,
  {\em Quasi-Normal Modes of Brane-Localised Standard Model Fields. II. Kerr Black Holes},
  Phys.\ Rev.\ D {\bf 74} (2006) 064008
  [gr-qc/0607048].
  

\bibitem{Nozawa:2008wf} 
  M.~Nozawa and T.~Kobayashi,
  {\em Quasinormal modes of black holes localized on the Randall-Sundrum 2-brane},
  Phys.\ Rev.\ D {\bf 78}, 064006 (2008)
  [arXiv:0803.3317 [hep-th]].


\bibitem{Aros:2002te}
  R.~Aros, C.~Martinez, R.~Troncoso and J.~Zanelli,
 {\em Quasinormal modes for massless topological black holes},
  Phys.\ Rev.\ D {\bf 67} (2003) 044014
  [hep-th/0211024].

\bibitem{Oliva:2010xn}
  J.~Oliva and R.~Troncoso,
 {\em Exact quasinormal modes for a special class of black holes},
  Phys.\ Rev.\ D {\bf 82} (2010) 027502
  [arXiv:1003.2256 [hep-th]].



\bibitem{Motl:2002hd}
  L.~Motl,
  {\em An Analytical computation of asymptotic Schwarzschild quasinormal frequencies},
  Adv.\ Theor.\ Math.\ Phys.\  {\bf 6} (2003) 1135
  [gr-qc/0212096].
  
\bibitem{Motl:2003cd}
  L.~Motl and A.~Neitzke,
 {\em Asymptotic black hole quasinormal frequencies},
  Adv.\ Theor.\ Math.\ Phys.\  {\bf 7} (2003) 307
  [hep-th/0301173].
  
\bibitem{Castro:2013lba}
  A.~Castro, J.~M.~Lapan, A.~Maloney and M.~J.~Rodriguez,
  {\em Black Hole Scattering from Monodromy},
  Class.\ Quant.\ Grav.\  {\bf 30} (2013) 165005
  [arXiv:1304.3781 [hep-th]].


\bibitem{Castro:2013kea} 
  A.~Castro, J.~M.~Lapan, A.~Maloney and M.~J.~Rodriguez,
  {\em Black Hole Monodromy and Conformal Field Theory},
  Phys.\ Rev.\ D {\bf 88}, 044003 (2013)
  [arXiv:1303.0759 [hep-th]].


\bibitem{Gonzalez:2010vv} 
  P.~Gonzalez, E.~Papantonopoulos and J.~Saavedra,
  JHEP {\bf 1008}, 050 (2010)
  [arXiv:1003.1381 [hep-th]].
  
  
\bibitem{Cardoso:2006nh} 
  A.~Cardoso, K.~Koyama, A.~Mennim, S.~S.~Seahra and D.~Wands,
  Phys.\ Rev.\ D {\bf 75}, 084002 (2007)
  [hep-th/0612202].
  
\bibitem{Cvetic:2014ina} 
  M.~Cvetic, G.~W.~Gibbons and Z.~H.~Saleem,
  Phys.\ Rev.\ D {\bf 90}, no. 12, 124046 (2014)
  [arXiv:1401.0544 [hep-th]].
  
 

\bibitem{QNMRev1} 
E.~Berti,  V.~Cardoso, and A.~O.~Starinets, {\em {Quasinormal modes of black holes and black branes}},
\newblock Class. Quant. Grav. {\bf 26}, 163001, 2009 [arXiv:0905.2975].

\bibitem{Konoplya:2011qq}
  R.~A.~Konoplya and A.~Zhidenko,
  {\em Quasinormal modes of black holes: From astrophysics to string theory},
  Rev.\ Mod.\ Phys.\  {\bf 83} (2011) 793
  [arXiv:1102.4014 [gr-qc]].



\bibitem{Plebanski:1976gy} 
  J.~F.~Plebanski and M.~Demianski,
  ``Rotating, charged, and uniformly accelerating mass in general relativity,''
  Annals Phys.\  {\bf 98}, 98 (1976).



\bibitem{Anber:2008qu}
  M.~Anber and L.~Sorbo,
  {\em New exact solutions on the Randall-Sundrum 2-brane: lumps of dark radiation and accelerated black holes},
  JHEP {\bf 0807} (2008) 098
  [arXiv:0803.2242 [hep-th]].
  
  
  
\bibitem{Chamblin:1999by}
  A.~Chamblin, S.~W.~Hawking and H.~S.~Reall,
  {\em Brane world black holes},
  Phys.\ Rev.\ D {\bf 61} (2000) 065007
  [hep-th/9909205].
  
  
  
\bibitem{Gregory:1993vy}
  R.~Gregory and R.~Laflamme,
 {\em Black strings and p-branes are unstable},
  Phys.\ Rev.\ Lett.\  {\bf 70} (1993) 2837
  [hep-th/9301052].
  
\bibitem{Gregory:2000gf}
  R.~Gregory,
  {\em Black string instabilities in Anti-de Sitter space},
  Class.\ Quant.\ Grav.\  {\bf 17} (2000) L125
  [hep-th/0004101].



\bibitem{Liu:2008ds} 
  L.~h.~Liu and B.~Wang,
  {\em Stability of BTZ black strings},
  Phys.\ Rev.\ D {\bf 78}, 064001 (2008)
  [arXiv:0803.0455 [hep-th]].


  
\bibitem{Zhidenko:2003wq} 
  A.~Zhidenko,
  {\em Quasinormal modes of Schwarzschild de Sitter black holes},
  Class.\ Quant.\ Grav.\  {\bf 21}, 273 (2004)
  [gr-qc/0307012].


  

\bibitem{Crispino:2013pya} 
  L.~�s C.~B.~Crispino, A.~Higuchi, E.~S.~Oliveira and J.~V.~Rocha,
 {\em Greybody factors for nonminimally coupled scalar fields in Schwarzschild�de Sitter spacetime},
  Phys.\ Rev.\ D {\bf 87}, no. 10, 104034 (2013)
  [arXiv:1304.0467 [gr-qc]].


  

  

 \end{thebibliography}
\end{document}